\documentclass[sn]{sn-jnl}

\usepackage{graphicx}%
\usepackage{multirow}%
\usepackage{amsmath,amssymb,amsfonts}%
\usepackage{amsthm}%
\usepackage{mathrsfs}%
\usepackage[title]{appendix}%
\usepackage{xcolor}%
\usepackage{textcomp}%
\usepackage{manyfoot}%
\usepackage{booktabs}%
\usepackage{algorithm}%
\usepackage{algorithmicx}%
\usepackage{algpseudocode}%
\usepackage{listings}%

\usepackage{xurl}
\usepackage{hyperref}
\hypersetup{hypertex=true,
	colorlinks=true,
	linkcolor=blue,
	anchorcolor=blue,
    urlcolor=blue,
    filecolor=mycustompurple,
	citecolor=blue}
 
\usepackage[noabbrev,nameinlink]{cleveref}
\crefname{figure}{Fig.}{Figures}
\crefname{table}{Table}{Table}

\usepackage{times}
\usepackage{graphicx}
\usepackage{utfsym}
\usepackage{fontawesome}

\usepackage{caption}
\begin{document} 

\title[Article Title]{A Neural-Network-Based Mapping and Optimization Framework for High-Precision Coarse-Grained Simulation} 

\author[1,2]{\fnm{Zhixuan Zhong}} 

\author[1,2]{\fnm{Lifeng Xu}} 

\author*[1,2]{\fnm{Jian Jiang}}\email{jiangj@iccas.ac.cn}

\affil[1]{\orgdiv{Beijing National Laboratory for Molecular Sciences, State Key Laboratory of Polymer Physics and Chemistry, Institute of Chemistry}, \orgname{Chinese Academy of Sciences}, \orgaddress{\city{Beijing}, \postcode{100190}, \country{P. R. China}}}

\affil[2]{\orgname{University of Chinese Academy of Sciences}, \orgaddress{\city{Beijing}, \postcode{100049}, \country{P. R. China}}}



\abstract{The accuracy and efficiency of a coarse-grained (CG) force field are pivotal for high-precision molecular simulations of large systems with complex molecules. We present an automated mapping and optimization framework for molecular simulation (AMOFMS), which is designed to streamline and improve the force field optimization process. It features a neural-network-based mapping function, DSGPM-TP (Deep Supervised Graph Partitioning Model with Type Prediction). This model can accurately and efficiently convert atomistic structures to CG mappings, reducing the need for manual intervention. By integrating bottom-up and top-down methodologies, AMOFMS allows users to freely combine these approaches or use them independently as optimization targets. Moreover, users can select and combine different optimizers to meet their specific mission. With its parallel optimizer, AMOFMS significantly accelerates the optimization process, reducing the time required to achieve optimal results. Successful applications of AMOFMS include parameter optimizations for systems such as POPC and PEO, demonstrating its robustness and effectiveness. Overall, AMOFMS provides a general and flexible framework for the automated development of high-precision CG force fields.}


\maketitle

\section*{Introduction}

In molecular science, accurate molecular characterization and analysis are crucial for elucidating the mechanisms within complex systems across various fields, including material synthesis \cite{Guo2023, Chen2024, Luke2023} and biological processes. \cite{Mao2024, Cayron2024, Darby2023} While experimental methods have their limitations in temporal and spatial resolution, coarse-grained molecular dynamics simulations (CGMD) is a powerful and efficient method, enabling extended investigations of the microscopic mechanisms of macroscopic systems.\cite{Noid2022, Borges-Araujo2023} In addition, coarse-grained (CG) molecular modeling provides a promising approach to bridge the atomic scale to micro scale, particularly in addressing the inherent limitations of all-atom models such as high computational cost and limited time scale. This method simplify the molecular complexity while retaining essential interactions and dynamics crucial for understanding the physics behind phenomena such as the self-assembly behaviors of amphiphiles.\cite{VanTeijlingen2023, Coscia2023, Parekh2023} 

Nevertheless, the accuracy of CGMD is critically dependent on the precise parameterization of the force field.\cite{shi2023, Zhu2022} Top-down and bottom-up approaches are the most common strategy for constructing or optimizing force field parameters. The top-down method focuses on fitting macroscopic properties, ensuring that the simulations align with observable phenomena at larger scales.\cite{Ingolfsson2014} Conversely, the bottom-up approach is concerned with reproducing microscopic statistics to maintain fidelity to atomic-level interactions and dynamics.\cite{jin2022} Among the universal CG force fields, the MARTINI force field \cite{Marrink2007, Souza2021} stands out due to its straightforward mapping scheme and effective partitioning of CG bead energies. These make it one of the most popular molecular force fields for CGMD, balancing simplicity with a robust capacity to model a wide range of molecular systems effectively. Despite its widespread use, achieving high accuracy and correctly reproducing microscopic physics across various systems remains a significant challenge for both MARTINI2 and MARTINI3 iterations. \cite{jin2022, Alessandri2019, Majumder2021, Jarin2021} In contrast, more specialized methods like the multiscale coarse-graining (MS-CG) technique \cite{Izvekov2005, Izvekov2005-2, Izvekov2006} adopt a bottom-up strategy, focusing on accurately reproducing local statistical correlations and atomic forces, thereby offering a more detailed and precise approach to parameterization that aligns closely with empirical data. Recently, deep learning offers an innovative and precise approach to develop bottom-up CG potentials, \cite{Friederich2021} e.g., the CGnets \cite{Wang2019} and neural network potentials (NNPs) for proteins \cite{Majewski2023}. Notably, the precision of these bottom-up CG force fields such as MS-CG series and CGnets heavily depends on the quality and detail of fine-grained datasets, such as atomistic trajectories and ab initio force/energy information. Moreover, bottom-up modeling approximates coarse-graining, reproducing structures at specific states. However,  this treatment often fails in transferability and representability across different states and thermodynamic properties. \cite{Dunn2016}

Additionally, the multiscale mapping process often requires meticulous manual intervention and expertise. Striking the right balance between atomic detail and computational efficiency without compromising the model's accuracy is a significant challenge. \cite{jin2022, Mancardi2023} The precise mapping of atomic features to CG beads and the accurate parameterization of the force field are crucial factors that impact the performance and reliability of CGMD simulations. In fact, there are several tools available for parameter optimization in CGMD. However, as shown in \cref{table1}, the existing algorithms for coarse-grained force field parameter optimization do not support automated mapping or the functionality for user-defined optimization schemes. To address these challenges, we present an automated mapping and optimization framework for molecular simulation (AMOFMS), a Python package, designed to streamline the mapping  and parameterization of CG force fields. This framework employs a graph-based neural networks to automate the mapping of atomistic structures to their corresponding CG representations. Moreover, AMOFMS incorporates multiple parallel optimizers, enabling more efficient exploration of complex molecular systems. 

\begin{table}[htbp]
     \centering
     
	\renewcommand\arraystretch{1.5}
	\caption{Comparison with current CG parameter optimization toolkit.}
	\label{table1}
	\begin{tabular}{ccccccc}
		\hline
		Toolkit & Year & Mapping & Top-down & Bottom-up & Optimization algorithm \\
		\hline
		VOTCA\cite{Ruhle2009}  & 2009 &  user-defined & \faTimes & \faCheck & IBI, IMC, FM \\
            MagiC\cite{Mirzoev2013} & 2013 & user-defined & \faTimes & \faCheck & IBI, IMC\\
		CAROL\cite{McDonagh2019}  & 2019 &  user-defined & \faCheck & \faTimes & Bayesian \\
		SwarmCG\cite{Empereur-Mot2020, Empereur-Mot2022} & 2020 & user-defined & \faCheck & \faCheck & FST-PSO \\
            coarsen\cite{Mahajan2023}  & 2023 & user-defined & \faTimes & \faCheck & Adaptive gradient descent \\
		OpenMSCG\cite{Peng2023}  & 2023 & user-defined & \faTimes & \faCheck & IBI, FM, REM \\
            CGCompiler\cite{Stroh2023} & 2023 & user-defined & \faCheck & \faCheck & mv-PSO \\
            \textbf{AMOFMS (ours)} & \textbf{2024} & \textbf{auto/user-defined} & \faCheck & \faCheck & \textbf{PSO, GA, Simplex, Bayesian}\\
		\hline
	\end{tabular}
	
	\footnotetext{\scriptsize IBI: Iterative Boltzmann Inversion; IMC: Inverse Monte Carlo; FM: Force Matching; \\REM: Relative Entropy Minimization; PSO: Particle Swarm Optimization; GA: Genetic Algorithm.}
\end{table}

\section*{Results}
Coarse-graining mapping and parameter optimization are crucial components for obtaining accurate and reliable results in CGMD simulations. AMOFMS is a comprehensive Python package that tailored for highly efficient parameter optimization of CG force fields, allowing users to implement customized rules and algorithms based on their personalized demands. In the next section, we will delineate the workflow for parameter optimization using AMOFMS and elucidate the functions of its principal modules.

 
\subsection*{Workflow of Parameter Optimization}
\begin{figure}[htbp]
	\centering
	\includegraphics[scale=0.8]{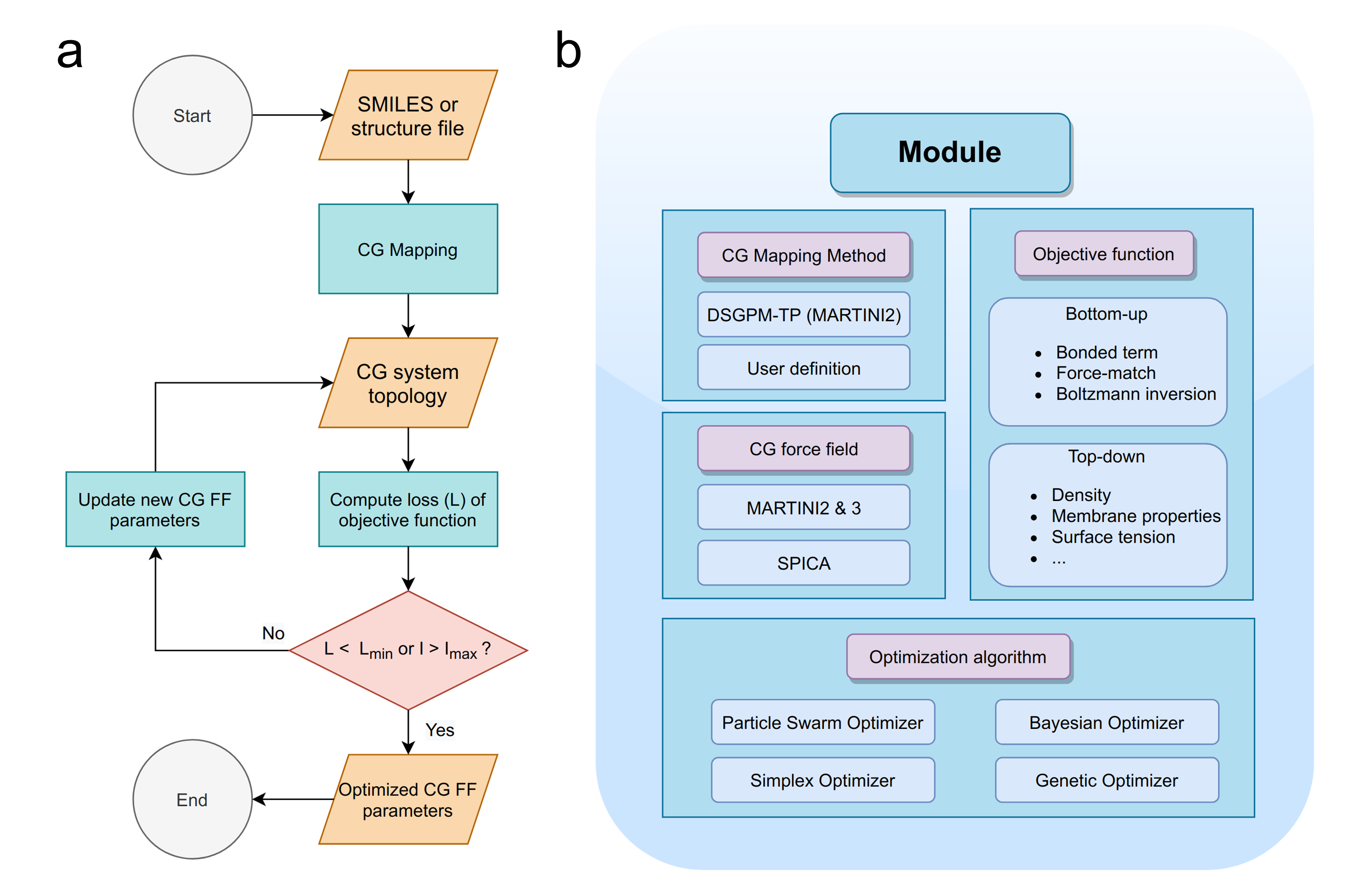}
	\caption{(a) Workflow of CG force field parameter optimization using AMOFMS; (b) modules of AMOFMS.}
	\label{Fig1}
\end{figure}

As illustrated in \cref{Fig1}a, AMOFMS initiates its workflow by accepting either SMILES or regular structure files (PDB, MOL2 and etc.) as input. Through a CG mapping process, these inputs are then transformed into a topological object in CG system,, which includes force field parameters. During this phase, users can employ the DSGPM-TP (Deep Supervised Graph Partitioning Model with Type Prediction Enhancement) algorithm, which based on the MARTINI mapping rules, or they can opt for a custom mapping definition to execute this process.

Subsequent to the CG mapping, AMOFMS computes the loss (L) of the objective function. The objective function is designed to quantify the loss associated with target properties, which may include top-down macroscopic properties such as density and surface tension, or bottom-up microscopic statistics like bond length and force, or potentially a combination of both.

Following the loss calculation, the optimizer evaluates whether the loss has reached a predefined minimum tolerance level (L$_ {\rm min}$) or if the current iteration has reached the specified maximum limit (I$_ {\rm max}$). If neither condition is met, the optimizer proceeds to update the CG force field parameters and repeats the cycle. This iterative process continues until the optimization criteria are satisfactorily met. Notably, our current AMFOMS exclusively supports parameter optimization for the MARTINI2,\cite{Marrink2007} MARTINI3,\cite{Souza2021} and SPICA (SDK)  \cite{Seo2019} force fields. Integration of user-defined potential functions is planned for future updates. The overall review of AMOFMS module is illustrated in \cref{Fig1}b.

\subsection*{Mapping Methods}
The process of coarse-grained mapping is an approach to reduce the number of degrees of freedom by grouping multiple atoms into larger units called ``CG beads". This reduction in complexity allows for more extensive and computationally feasible simulations. In general, selecting how atoms are grouped into beads relies heavily on the experts' experiences and chemical and physical intuition.\cite{Ingolfsson2014} For instance, the popular MARTINI CG model typically groups four heavy atoms into a single CG bead. \cite{Marrink2013} However, to improve the scalability and transferability of CG models, automating the mapping process is important. To address this, motivated from DSGPM (Deep Supervised Graph Partitioning Model) \cite{Li2020},  we propose a novel graph neural network framework named the DSGPM-TP (DSGPM with Type Prediction Enhancement), which can be used to automatically map atomistic structures to CG beads. This approach is aimed at replacing human experts' intuition or experience. The DSGPM-TP formulates the prediction of CG mappings as a binary task that includes both a graph partitioning problem (atom clustering) and a multi-class classification task (atom-to-CG type prediction). This dual approach allows the model to first determine clusters of atoms that should be grouped together into beads (graph cut), and then categorize each cluster into specific types of CG beads based on their chemical and physical properties (multi-class classification). 


\begin{figure}[htbp]
	\centering
	\includegraphics[scale=0.8]{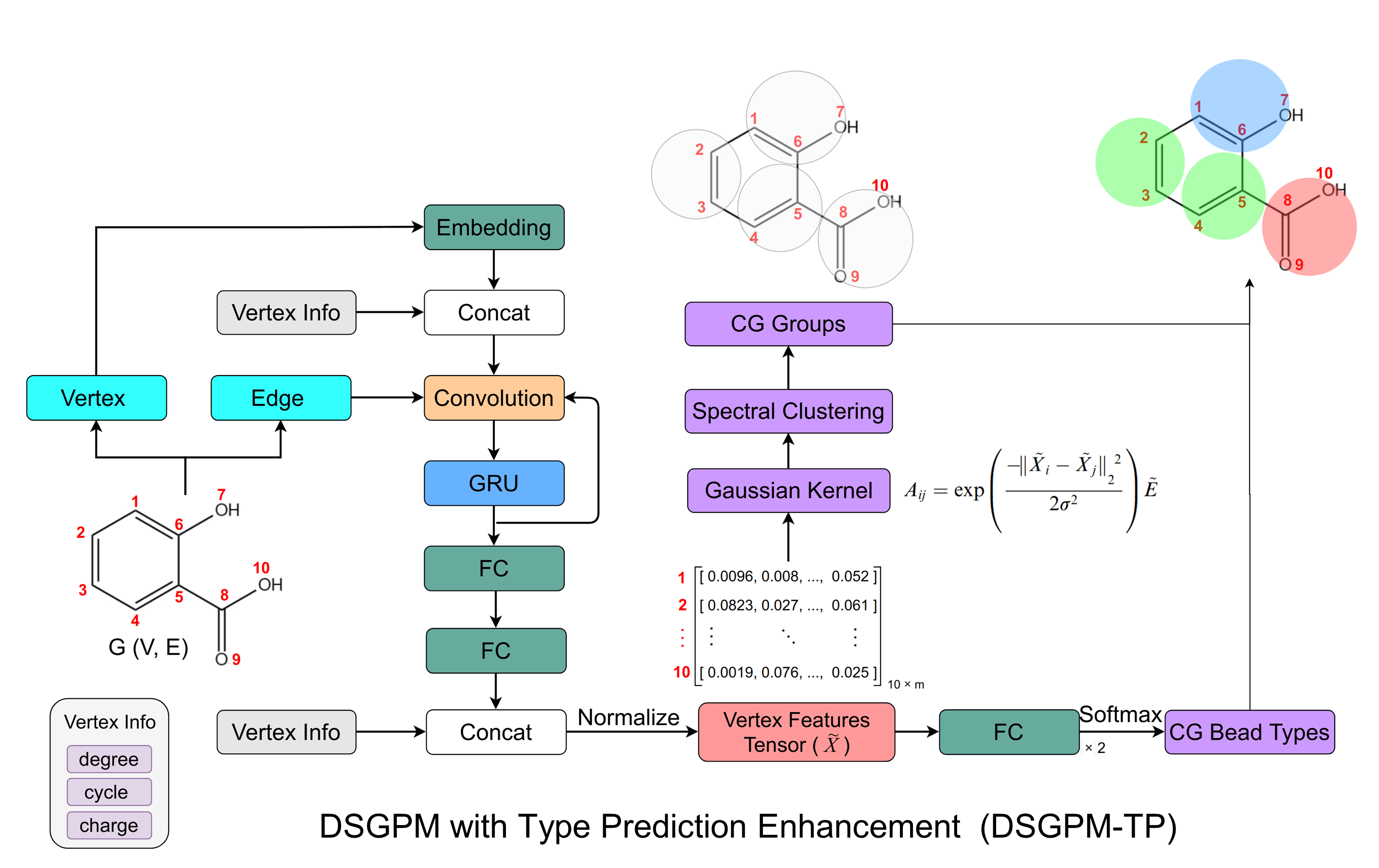}
	\caption{Architecture of DSGPM-TP. FC denotes full-connected layer. Concat is concatenation. GRU refers to gated recurrent units.}
	\label{Fig2}
\end{figure}

\begin{figure}[htbp]
	\centering
	\includegraphics[scale=0.8]{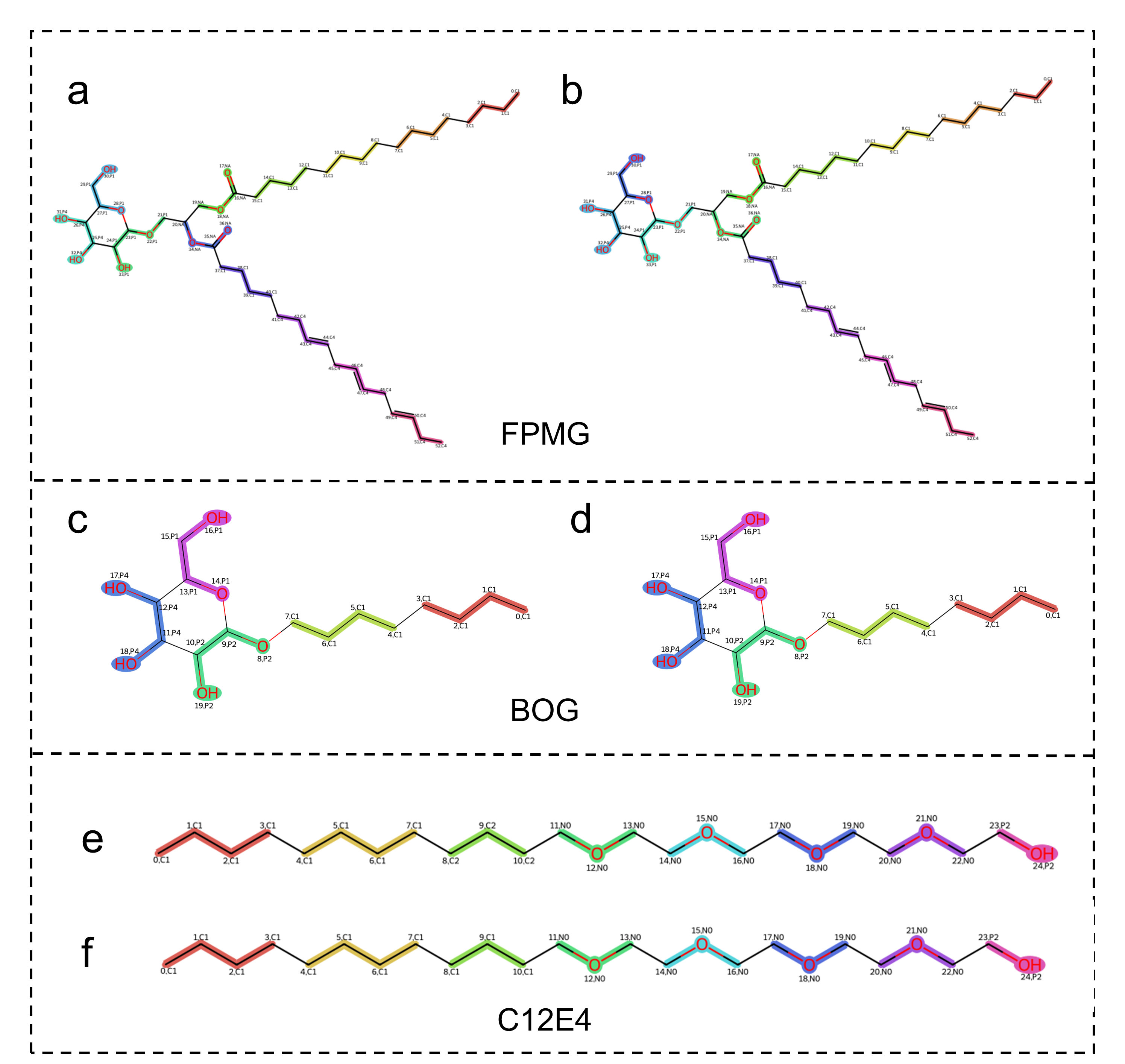}
	\caption{Visualization of the CG mapping prediction and the ground truth. (a), (c), (e) are the ground truth. (b), (d), (f) are the results predicted by DSGPM-TP model. Atoms and their corresponding edges that belong to the same CG bead are highlighted with the same color. FPMG: C18:3/16:0 monogalactosyldiacylglycerol; BOG: 2-(hydroxymethyl)-6-octoxyoxane-3,4,5-triol; C12E4: tetraethylene glycol monododecyl ether. }
	\label{Fig3}
\end{figure}

As shown in \cref{Fig2}, this architecture is designed to effectively generate robust atom features for spectral clustering and multi-class classification. It is based on the message passing neural network (MPNN) framework \cite{Gilmer2017}, enhancing its capability by concatenating specific atom metrics such as degree, charge, and cycle feat. This integration allows the model to capture the representative information of each atom, facilitating more accurate clustering and classification by providing comprehensive insights into the chemical environment and connectivity of the atoms. For the task of clustering atoms, as demonstrated in \cref{table2}, our method surpasses existing state-of-the-art methods in terms of both AMI (Adjusted Mutual Information) \cite{XuanVinh2010} and graph cut metrics. This result confirms the effectiveness of our model in accurately capturing the essential features of molecular structures. Subsequently, the CG mapping results against ground-truth are visualized in Fig. \ref{Fig3} and S1.  Overall, the DSGPM-TP shows good performance in CG mappings. Additionally, a series of ablation studies are conducted to examine the contributions of type loss,  non-cut pair loss, vertex features, and the number of fully-connected layers to the CG type prediction. The results, as shown in Table S1-S4, indicate the impact of each component on the overall performance of the DSGPM-TP model and their individual roles in enhancing prediction accuracy. Moreover, this comprehensive analysis validates the design choices made in developing DSGPM-TP.

While DSGPM-TP currently focuses on CG mappings based on MARTINI model, it can adapt to new consensus-based mappings in the future by training on updated datasets. This flexibility ensures DSGPM-TP remains a state-of-the-art tool for automating CG mapping.

\begin{table}[htbp]
    \centering
	\renewcommand\arraystretch{1.5}
	\caption{Comparison with state-of-the-art methods.}
	\label{table2}
	\begin{tabular}{ccccc}
		\hline
		Method & AMI & Cut Prec. & Cut Recall & Cut F1-score \\
		\hline
		HDBSCAN \cite{HDBSCAN} & 0.5117 & 0.4631 & 0.4857 & 0.4205 \\
		FINCH \cite{Sarfraz2019} & 0.6030 & 0.4602 & 0.8015 & 0.5595 \\
		h-NNE \cite{Sarfraz2022} & 0.4230 & 0.3390 & 0.7637 & 0.4510 \\
		Graclus \cite{Dhillon2007} & 0.4205 & 0.2614 & 0.6826 & 0.3701 \\
		METIS \cite{Karypis1998} & 0.7462 & 0.4968 & 0.5218 & 0.5038 \\
		Spectral Cluster \cite{Ng2001}  & 0.8375 & 0.7000 & 0.6945 & 0.6971  \\
		DSGPM \cite{Li2020} & 0.9139 & 0.8302 & 0.8314 & 0.8308 \\
 		\textbf{DSGPM-TP} \textbf{(ours)}  &  \textbf{0.9510} & \textbf{0.9029} &  \textbf{0.9049} &  \textbf{0.9038} \\
		\hline
	\end{tabular}
	
\end{table}


\subsection*{Objective Functions}
Parameter optimization is typically treated as a multi-objective optimization problem (MOP). This type of optimization involves managing several competing objectives that need to be simultaneously satisfied to achieve the most desirable overall outcome. A straightforward technique for implementing a MOP is scalarization, which involves converting the multiple objectives into a single objective. One common scalarization method is linear weighting, which is given by
\begin{equation}
    Loss = \sum \limits_{i} {w_i f_i(\textbf{x})} \ ,
\end{equation}
where $w_i$ represents the weight assigned to the 
$i$-th objective, $f_i$ denotes the $i$-th objective function, and $\textbf{x}$ is the vector of parameters being optimized. The objective weights $w_i$ play a crucial role in the optimization process as they help to balance the different parametrization targets. The setting of weight parameters is influenced by numerous factors, such as the specific molecular system, the requirements of the modeling task, and the quality of the training dataset. Using AMOFMS, users have the flexibility to configure multiple objective functions tailored specifically for the top-down process, the bottom-up process, or a combination of both. 

\subsection*{Optimizer}
In the AMOFMS, the optimizer module is responsible for calculating the objective functions corresponding to a given set of CG force field parameters and subsequently updating these parameters until the convergence criteria are met. Choosing an appropriate optimizer is a critical decision in the optimizing process, as it significantly impacts the efficiency and effectiveness of a optimization task.  In the current version of AMOFMS, four optimizers are involved: Particle Swarm Optimizer (PSO), Bayesian Optimizer (BO), Genetic Optimizer (GO), and Simplex Optimizer (SO). These four optimization algorithms have been extensively employed across various fields, particularly in the development of force fields.\cite{McDonagh2019, Stroh2023, Mishra2018, Chan2019, Weiel2021} In our package, all optimizers utilize a parallel architecture to accelerate computational processes, and users have the flexibility to select either a single optimizer or a combination of multiple optimizers to effectively address the optimization task. 

In the subsequent sections, we will illustrate the capabilities of AMOFMS through two practical examples, i.e., 1-palmitoyl-2-oleoylphosphatidylcholine (POPC) bilayer system with MARTINI2 Force Field and polyethylene oxide (PEO) systerm with MARTINI3 force field. Generally, we will outline the necessary steps to provide a concise overview of the workflow from initial setup to execution. The optimized parameters (denoted as Optimized CG) are then used in simulations and compared with the results from simulations based on the original force field parameters (denoted as Original CG) as well as experimental values to validate their accuracy and reliability. Each optimized and original system are carried out with three independent replicas, enabling us to ensure reproducibility and statistical reliability of the results.

\subsection*{Example 1: POPC with MARTINI2}

POPC is a widely studied phospholipid that forms bilayer membranes, serving as model systems for biological membranes. The structural and dynamical properties of POPC membranes are crucial for understanding membrane behavior and interactions in biological systems. In this section, we demonstrate the optimization process of MARTINI2 CG force field parameters for the POPC membrane and present the results from the optimized model.

\begin{figure}[htbp]
	\centering
	\includegraphics[scale=0.8]{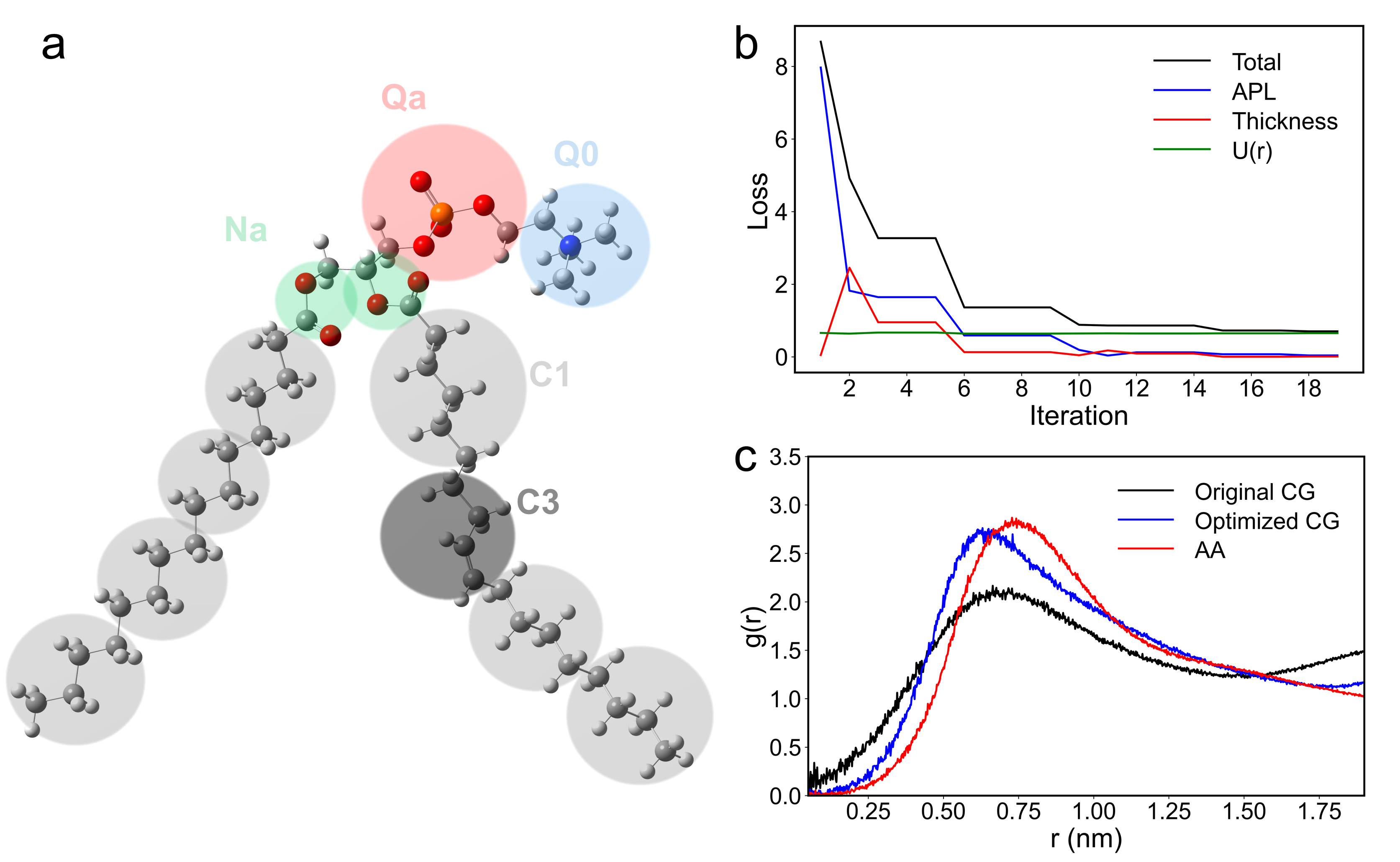}
	\caption{(a) CG mapping of POPC predicted by DSGPM-TP model; (b) loss as a function of iterations; (c) radial distribution function between POPC molecules.}
	\label{Fig4}
\end{figure}

The CG mapping of the POPC molecule based on the DSGPM-TP model 
proposed in this work is illustrated in \cref{Fig4}a. This mapping result aligns with the one recommended by the developers of the MARTINI force field, ensuring consistency and reliability in our automated mapping model. The goal of this task is to fit both potential of mean force (U(r)) and macroscopic properties including the area per lipid (APL) and membrane thickness. The total loss ($L$) can be expressed as a weighted sum of deviations from target values for these properties: 
\begin{equation}
    L = L_{U(r)} + 10 \times (L_{\rm APL} + L_{\rm thickness}) .
\end{equation}
Subsequently, the PSO optimizer with 32 swarms and 20 iterations is employed to execute the optimization process. As shown in \cref{Fig4}b, the total loss rapidly decreases with iterations. \cref{table3} compares the experimental values and those calculated by the original and optimized MARTINI2 models, demonstrating that the results from the optimized CG model aligns more closely with experimental data. Besides the APL and membrane thickness, the volume per lipid (VPL), isothermal area compressibility modulus (K$\rm _A$), and the lateral diffusion coefficient  (D$_L$) obtained from the optimized CG model also show good agreement with experiments, indicating that the optimizer not only accurately captures the structural properties of the POPC membrane but also reproduces other important properties such as the dynamic behavior (D$_L$). It is worth mentioning  that, to avoid the effects of finite size, similar comparisons are conducted in larger boxes, as shown in Table S5. Furthermore, the employed bottom-up strategy ensures that the radial distribution function g(r) between POPC molecules shows greater consistency with fine-grained models, such as the CHARMM all-atom force field (AA), as displayed in \cref{Fig4}c. Moreover, bond and angle distributions obtained from the optimized CG model also show better agreement with the all-atom models (Fig. S2).  These results demonstrates the capability of our AMOFMS to accurately develop  coarse-grained models for complex biological systems.

\begin{table}[htbp]
    \centering
	\renewcommand\arraystretch{1.5}
	\caption{Properties of POPC membrane.}
	\label{table3}
	\begin{tabular}{cccccc}
		\hline
		  Method & Thickness (nm) & APL (nm$^ {2}$)  & VPL(nm$^{3}$) & D$_L$ (*10$^{-8} \rm cm^2/s$) & K$\rm _A$ (mN/m) \\
		\hline
		Original CG& 4.019 $\pm$ 0.001 &  0.660 $\pm$ 0.000  & 1.372 $\pm$ 0.000 & 13.884 $\pm$ 0.365 & 306.612 $\pm$ 7.361\\
		\textbf{Optimized CG (ours)} & \textbf{3.724 $\pm$ 0.001} & \textbf{0.629 $\pm$ 0.002} & \textbf{1.203 $\pm$ 0.000} & \textbf{11.899 $\pm$ 0.360} & \textbf{309.980 $\pm$7.050} \\
		Experiment & 3.70\cite{Kucerka2006}  &  0.63\cite{Smaby1997} & 1.223\cite{Pabst2000} & 8.87-10.70 \cite{Filippov2003} & 180-330  \cite{Binder2001}\\
		\hline
	\end{tabular}
	
\end{table}

\subsection*{Example 2: PEO with MARTINI3}

PEO is a polymer of significant interest in various fields such as materials science, biomedicine, and nanotechnology. Optimizing the force field parameters for PEO is crucial to ensure accurate predictions of the structural, themodynamic, and dynamic properties of PEO systems. In this section, we outline the optimization process of the MARTINI3 CG force field parameters for PEO and discuss the results of the optimized model.

\begin{figure}[htbp]
	\centering
	\includegraphics[scale=0.8]{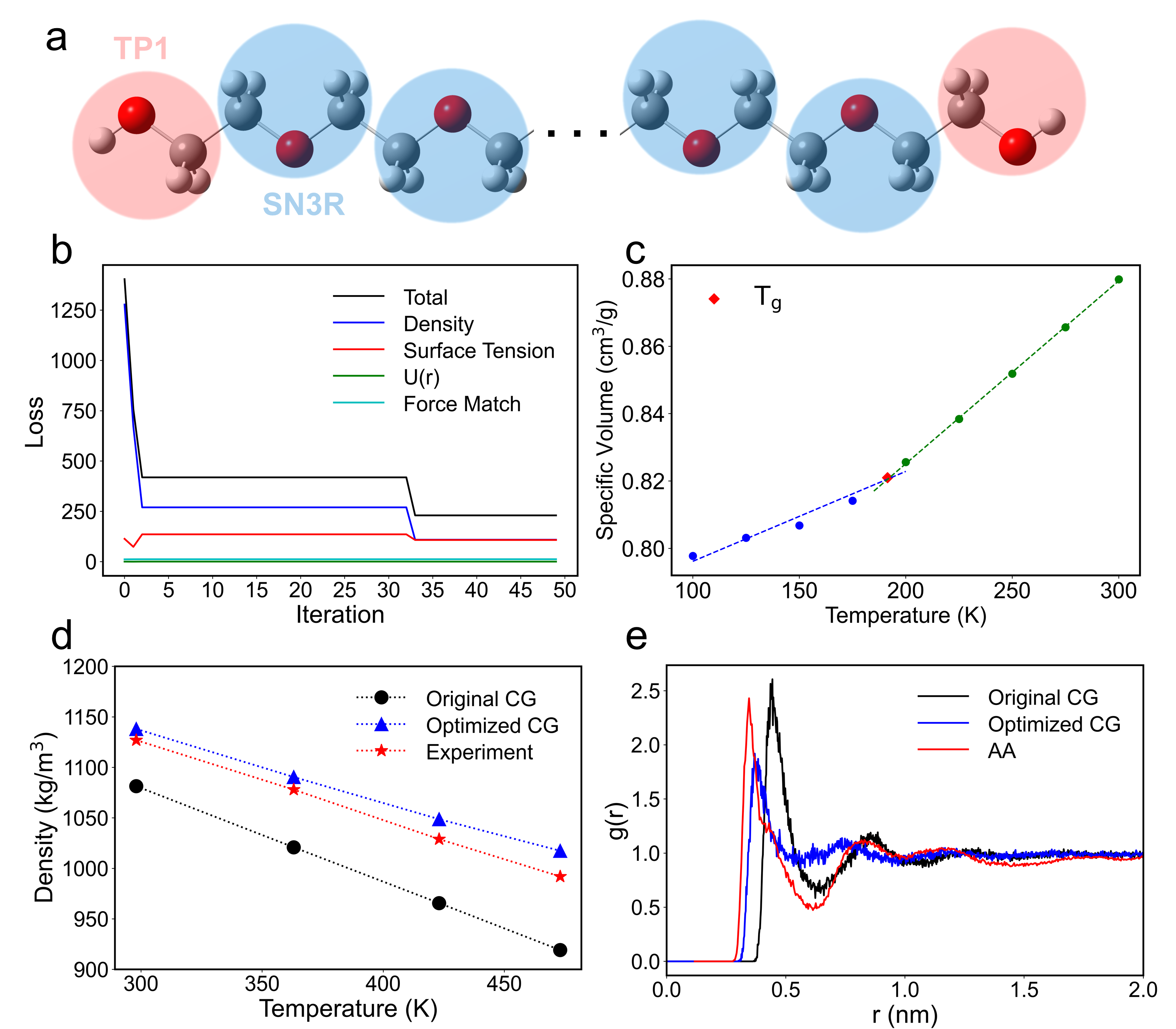}
	\caption{(a) CG mapping of PEO; (b) loss as a function of iterations; (c) prediction of the glass transition points of PEO; (d) density of PEO system under different temperatures; (e) radial distribution function between TP1 beads.}
	\label{Fig5}
\end{figure}

The PEO molecule is modeled based on the MARTINI3 representation, and the initial force field parameters are generated by the Polyply python software. \cite{Grunewald2022} It is important to note that we use the TP1 bead to represent the termini of the polymer chain, rather than the SN3R bead recommended by Polyply, as shown in the \cref{Fig5}a. This is because TP1 bead provides a more accurate polarity of the termini in the framework of the standard MARTINI3 representation. In this work, the system consists of 40 PEO chains with a polymerization degree of 52. The optimization process combines both bottom-up and top-down strategies. The bottom-up approach focuses on fitting the potential of mean force (U(r)) and CG forces, while the top-down approach targets macroscopic properties such as density and surface tension. This dual strategy ensures a comprehensive optimization that captures both micro- and macroscopic  properties. The total loss is defined as
\begin{equation}
    L = L_{U(r)} + L_{\rm FM} + 10 \times L_{\rm density} + 50 \times L_{\rm surface\ tension} .
\end{equation}
The Baysian Optimizer is employed to perform the optimization.  As shown in the \cref{Fig5}b, this optimizer demonstrates a robust ability to reduce the loss function efficiently, achieving this with a lower computational cost (50 iterations with 200 initial samples).

\begin{table}[htbp]
    \centering
	\renewcommand\arraystretch{1.5}
	\caption{Properties of PEO system.}
	\label{table4}
	\begin{tabular}{cccccc}
		\hline
		  Method & Density (kg/m$^3$) & Surface Tension (mN/m)  & T$\rm _g$ (K) \\
		\hline
		Original CG& 1081.430 $\pm$ 0.014 &  31.887 $\pm$ 0.482  & 188.700 \\
		\textbf{Optimized CG (ours)} & \textbf{1138.110 $\pm$ 0.093} & \textbf{40.077 $\pm$ 1.049} &  \textbf{191.400}\\
		Experiment & 1127 \cite{Wixwat1991}  &  42.6 \cite{Wixwat1991} & 201.15 \cite{Singla2008}\\
		\hline
	\end{tabular}
	
\end{table}
Although the original CG force field performed good, our AMOFMS-optimized CG force field shows even better agreement with experimental results, as shown in \cref{table4}, where the glass transition temperature is identified by observing the slope of the curve of the temperature-dependent specific volume (\cref{Fig5}c). Moreover, when applied to different temperatures, our CG force field predicts densities with errors within 1\% compared to experiments (\cref{Fig5}d), further demonstrating the success of AMOFMS. We have to note that, as shown in Table S6, the simulation results mentioned above are not affected by the size of the simulation box. Meanwhile, according to Fig. \ref{Fig5}e and  S3, our CG force field exhibits improved microscopic structure of PEO.

In summary, the successful applications of AMOFMS in both the POPC membrane and bulk PEO systems demonstrate its versatility and effectiveness. In fact, our AMOFMS can be utilized in a variety of systems to accurately study their microscopic and macroscopic properties.

\section*{Discussion}
In this study, we have presented a comprehensive tool, AMOFMS (an Automated Mapping and Optimization Framework for Molecular Simulation), designed to streamline and enhance the development and optimization of coarse-grained (CG) force fields. AMOFMS addresses several critical challenges in the field of molecular simulations such as the automated CG mapping and flexible optimization framework, offering significant advancements over existing methodologies. Some of the key contributions of AMOFMS are listed as below:

\textbf{1. Automated mapping functionality:} One of the most notable features of AMOFMS is its automated mapping capability. This functionality simplifies the complex and labor-intensive process of CG mapping, thereby reducing human error and improving the reproducibility of molecular simulations.

\textbf{2. Flexible optimization framework:} AMOFMS allows users to define and customize a wide range of optimization parameters, e.g., Lennard-Jones potentials, bond lengths, force constants and so on. This flexibility enables the tool to be applied to a variety of molecular systems from simple systems with small organic molecules to complex systems with biomacromolecules.

\textbf{3. Multi-objective optimization:} AMOFMS supports multi-objective optimization by integrating both bottom-up (e.g., pair distribution function and force matching) and top-down (e.g., density and surface tension) approaches. This comprehensive framework ensures that the force fields trained by our AMOFMS can simultaneously account for both the microscopic and macroscopic properties of the system.

\textbf{4. Parallel optimizer:} The parallel optimizers greatly accelerate the optimization process, enabling AMOFMS to efficiently handle large and complex systems.

Despite its the numerous advantages of AMOFMS, there are still some limitations that need to be addressed in future work:

1. Challenges in directly incorporating many-body interactions: Similar to most molecular force fields, AMOFMS considers intermolecular forces based on pairwise potential assumptions and does not directly account for many-body interactions. Accurately incorporating many-body interactions remains a significant challenge in molecular simulations.

2. Limit to specific force field models: In its current version, AMOFMS trains force fields using predefined models like the MARTINI models. This limitation may restrict its applicability to specialized molecular systems.

3. Underdeveloped gradient descent optimizer: The gradient descent optimizer has yet to be fully developed. However, deriving the analytical solutions for the differential forms of macroscopic properties with respect to force field parameters present significant challenges in our current version of AMOFMS.  In fact, an automatic differentiation technique may offer a promising solution to this problem. One such example is DMFF, an open-source molecular force field development platform based on automatic differentiation.\cite{Wang2023} We will consider automatic differentiation technique in our future work.\\

In conclusion, AMOFMS offers a robust, accurate, and efficient solution for development and optimization of force field. Its excellent features including  automated mapping, flexible parameter selection, multi-objective optimization, and parallel optimizer collectively provide a powerful toolkit for force field development. Thus, it enables more accurate and efficient computational simulations, advancing our understanding of complex molecular systems. Furthermore, although the current version of AMOFMS is limited to specific force field models, its framework shows promising potential for the development and optimization of other fine- or coarse-grained force fields.

\section*{Methods}
\subsection*{DSGPM-TP}
Deep Supervised Graph Partitioning Model with Type Prediction Enhancement (DSGPM-TP) treats the CG mapping prediction as a graph partitioning problem. In this model, a molecule is represented as a graph $G = (V, E)$, where $V$ consists of atom types encoded as one-hot vectors and $E$ encodes bond types. One motivation for DSGPM-TP is to enhance spectral clustering, which heavily depends on the quality of an affinity matrix $A$. Spectral clustering performs well when the affinity matrix accurately represents the relationships between nodes. However,  traditional methods often fall short by merely indicating the presence or absence of bonds but without capturing bond types or strengths. DSGPM-TP aims to create a refined affinity matrix that accurately distinguishes between different types of atomic interactions. Another motivation is to predict the type of the CG bead (cluster). This involves precisely identifying which atoms belong to the same CG bead type. This allows the model to provide insights into the global information of the molecule, which is crucial for practical applications across various fields such as molecular design.

\subsubsection*{Details of Model}
Based on the MPNN framework, the graph neural network (GNN) extracts features for each atom, initially projecting one-hot encoded atom types into a higher-dimensional feature space through a multilayer perceptron (MLP). As shown in \cref{Fig2}, this embedded feature is then enhanced with the atom's degree (the number of chemical bond directly connected to the atom), charge, and a cycle indicator, resulting in a vertex feature tensor $\tilde{X}$. Specifically, the features generated by the GNN are updated using convolutional layers and gated recurrent units (GRU) to capture complex relationships between atoms. The feature matrix ($ X_0$) is updated iteratively using the GRU layer over $T$ time steps to produce the final feature matrix $X_T$.
\begin{equation}
\begin{gathered}
\hat{X}_u{ }^{t-1}=\mathbf{W}^{\prime} X_u{ }^{t-1}+\sum_{v \in \mathcal{N}(u)} X_v{ }^{t-1} \phi^e\left(E_{u v}\right), \\
X_i{ }^t=\operatorname{GRU}\left(\hat{X}_u{ }^{t-1}, H_u{ }^{t-1}\right),
\end{gathered}
\end{equation}
where underscript $u$ denotes $u$-th atom and superscript $t$ denotes time step (layer); $\mathbf{W}^{\prime}$ is a weight matrix;  superscript $\prime$ denotes transpose; the function $\phi^e(\cdot)$ is used to map the bond type $E_{uv}$ to an edge-conditioned weight matrix, which is implemented as a MLP; and $H_u^{t-1}$ is the hidden state of the GRU for atom $u$ at the previous time step $t-1$. Subsequently, vertex features tensor $\tilde{X}$ is obtained by:
\begin{equation}
\begin{gathered}
\tilde{X}^{\prime}=\operatorname{Concat}\left(\operatorname{MLP}\left(X^T\right), V, I_{\mathrm{degree}}, I_{\mathrm{charge}}, I_{\mathrm{cycle}}\right), \\
\tilde{X}=\frac{\tilde{X}^{\prime}}{\left\|\tilde{X}^{\prime}\right\|_2},
\end{gathered}
\end{equation}
where $I_{\mathrm{degree}}$, $I_{\mathrm{charge}}$, and $I_{\mathrm{cycle}}$ represent degree, charge, and cycle indicator of each atom in a molecular graph, respectively. Here, the charge information is computed by Gasteiger-Marsili method implemented on RDKIT.\cite{Gasteiger1980}
Then, the affinity matrix \( A \) is computed using a Gaussian kernel, which measures the similarity between atom features:
\begin{equation}
    A_{ij} = \exp \left( -\frac{\| \tilde{X}_i - \tilde{X}_j \|^2}{2\sigma^2} \right) \tilde{E},
\end{equation}
where $\sigma=1$ is the bandwidth. $\tilde{E}$ represent the adjacency matrix:
\begin{equation}
\tilde{E}_{ij} = 
\begin{cases} 
1, & \text{if atom } i \text{ and atom } j \text{ are bonded} \\
0, & \text{otherwise}
\end{cases}
\end{equation}

\subsubsection*{Training}
To train the model, the total loss ($L$) consist of graph cut loss ($L_{\text {cut}}$) and the type prediction loss ($L_{\text {type}}$):
\begin{equation}
L = L_{\text {cut}}+\lambda_1 L_{\text {type}} .
\end{equation}
Graph cut loss is a type of loss function used in graph-based segmentation tasks. It measures the dissimilarity between regions in a graph that have been partitioned or segmented. Two types of cut losses are used: cut triplet loss ($L_{\text {triplet }}$) and non-cut pair loss ($L_{\text {pair }}$). The cut triplet loss encourages the model to separate features of atoms in different partitions by a margin ($\alpha$): 
\begin{equation}
L_{\text {triplet }} =\frac{1}{|N_T|} \sum_{\substack{T_i=\{\mathrm{a}, \mathrm{p}, \mathrm{n}\}}} \max \left(\left\|\tilde{V}_{\mathrm{a}}-\tilde{V}_{\mathrm{p}}\right\|_2-\left\|\tilde{V}_{\mathrm{a}}-\tilde{V}_{\mathrm{n}}\right\|_2+\alpha, 0\right) ,
\end{equation}
where \(T_i=\{\mathrm{a}, \mathrm{p}, \mathrm{n}\}\) denotes a specific triplet consisting of an anchor atom (\(\mathrm{a}\)), a positive atom (\(\mathrm{p}\)), and a negative atom (\(\mathrm{n}\)). \(\tilde{V}_{\mathrm{a}}\), \(\tilde{V}_{\mathrm{p}}\), \(\tilde{V}_{\mathrm{n}}\) are the feature vectors (embeddings) of the anchor, positive, and negative atoms, respectively.
On the other hand, the non-cut pair loss unifies the features of atoms within the same partition.
\begin{equation}
L_{\text {pair }}  =\frac{1}{|N_P|}\left\|\tilde{V}_{\mathrm{a}}-\tilde{V}_{\mathrm{a}^{\prime}}\right\|_2 .
\end{equation}
The final graph cut loss is defined as
\begin{equation}
L_{\text {cut }}  =L_{\text {triplet }}+\lambda _2 L_{\text {pair }} .
\end{equation}
Type prediction loss is a loss function used to train models for predicting the type for the coarsed-grained group or cluster. The cross-entropy loss function is commonly used for this classification task, i.e.,
\begin{equation}
L_{\text {type}} =-\log \left(\frac{\text{exp}({V_{\text{real}})}}{\sum_j \text{exp}({V_j})}\right) .
\end{equation}

The DSGPM-TP model is trained for 500 epochs, and the epoch that yields the best performance during the 5-fold cross-validation is selected. The hidden feature dimension is set to 128. The GNN implementation is built using PyTorch and PyTorch Geometric.\cite{NEURIPS2019_bdbca288, Fey_Lenssen_2019}

\subsubsection*{Dataset}
The MARTINI2 dataset used to train and validate models in this work is collected from the published literature and MARTINI website (\url{http://md.chem.rug.nl/index.php/example-applications2}), encompassing a wide range of molecular structures and properties documented in scientific research. Each entry in the dataset comprises information regarding the molecular coarse-grained (CG) groups, e.g., the type of each group and the details about which atoms belong to this group. The current dataset consists of 744 entries.

\subsection*{MD settings and analysis of POPC Example}
In this work, all the simulations (except for the analyses of surface tension) are performed in the isothermal-isobaric (NPT) ensemble using GROMACS 2021.4,\cite{Lindahl2021} with periodic boundary conditions applied in all three spatial directions.

\subsubsection*{Details of fine-grained simulation}
The initial POPC bilayer systems are constructed by CHARMM-GUI server.\cite{Jo2008}  Each system is composed of 384 POPC molecules and 14592 water molecules (box size $\sim \rm 10.75\ nm \times 10.75\ nm \times 8.04\ nm$). The CHARMM36 all-atom force field describes POPC molecules \cite{Klauda2010} and the TIP3P model is used for water molecules.\cite{Jorgensen1983} The van der Waals interactions are modeled using the Lennard-Jones potential with a cutoff radius of 1.2 nm. Electrostatic interactions are calculated with the particle mesh Ewald method, also employing a 1.2 nm cutoff.\cite{Essmann1995} Initially, the steepest descent algorithm is utilized to remove any unreasonable structures in the initial configuration.\cite{Wardi1988} The target temperature and pressure for all NPT simulations are set to 300 K and 1.01325 bar, respectively. Subsequently, a 100 ns simulation is conducted to ensure the system achieves equilibrium. The Bussi–Donadio–Parrinello thermostat \cite{Bussi2007} and the Berendsen semi-isotropic barostat are employed.\cite{ Berendsen1984} An additional 100 ns simulation is then performed for sampling, maintaining the temperature and pressure using the Nosé–Hoover thermostat \cite{Nose1984, Hoover1985} and the Parrinello–Rahman semi-isotropic barostat, respectively.\cite{Parrinello1981, Nose1983} Throughout the simulations, the LINCS algorithm \cite{Hess1997} is employed to constrain all bond lengths including hydrogen atoms, allowing for a larger integration time step, i.e., 2 fs.

\subsubsection*{Details of coarse-grained simulation}
PACKMOL software \cite{Martinez2009} and CHARMM-GUI webserver\cite{Jo2008} are utilized to construct the initial bilayer systems, each composing of 384 POPC molecules and 3649 water beads (box size $\sim \rm 11.02\ nm \times 11.02\ nm \times 7.67 \ nm$). To consider the finite size effect on the optimized parameters, two systems of 21 nm $\times$ 21 nm $\times$ 17 nm and 31 $\times$ 31 nm $\times$ 27 nm are also investigated. The MARTINI2 force field \cite{Marrink2007} is employed in these simulations. Notably, 10\% of the original water beads (P4 type) are replaced with anti-freezing water beads (BP4 type) to prevent freezing effects at lower temperatures and enhance system stability. The basic simulation settings are nearly identical to those used in the fine-grained simulations, with a few exceptions. A cutoff of 1.1 nm is chosen for nonbonded interactions to accelerate the simulations. The Bussi–Donadio–Parrinello thermostat \cite{Bussi2007} is selected for both equilibrium and production phases. The simulation time for equilibrium and production phases are 500 ns with a time step of 20.0 fs used throughout the simulations.

\subsubsection*{Analyses}
The thickness and area per lipid (APL) are computed by LiPyphilic \cite{Smith2021} using Freud,\cite{Ramasubramani2020} and the volume per lipid (VPL) is calculated by OVITO.\cite{Stukowski2010}

The lateral diffusion coefficient ($D\rm _L$) of lipids in a bilayer is obtained via the Einstein relation:
\begin{equation}
D_{\rm L}=\frac{1}{4} \lim _{t \rightarrow \infty} \frac{d}{d t}\left\langle\frac{1}{N} \sum_{i=1}^N\left|r_i\left(t_0+\Delta t\right)-r_i\left(t_0\right)\right|^2\right\rangle_{t_0},
\end{equation}
where \( N \) is the number of lipids, \( r_i(t_0) \) is the position in the \( xy \)-plane of lipid \( i \) at a origin time \( t_0 \), \( r_i(t_0 + \Delta t) \) is the position at a lag time \( \Delta t \), and the angular brackets denote an average over all origin times, \( t_0 \). It is worth noting that the Martini dynamics are faster than all-atom dynamics due to smoother coarse-grained interactions and the absence of fine-grained friction. A standard conversion factor of 4 is used for interpreting simulation results, reflecting the speed-up in diffusion dynamics of Martini water compared to real water. Therefore, the diffusion coefficient value should be corrected by dividing them by 4.\cite{Marrink2007}


The isothermal area compressibility modulus (K$_{\rm A}$) is computed from the fluctuation of the APL via 
\begin{equation}
\rm K_{A} = \frac{2k_B T \left\langle APL \right\rangle}{N {\delta^{2}_{APL}} },
\end{equation}
where $\rm  k_B$ is Boltzmann constant, T is the temperature, $\rm  \left\langle APL \right\rangle $ is the average of APL, $N$ is the total number of lipid and ${\rm  \delta^{2}_{APL}}$ is the variance of APL.

\subsection*{MD settings and analysis of PEO Example}
\subsubsection*{Details of fine-grained simulation}
PACKMOL software \cite{Martinez2009} is utilized to construct the initial systems, each composing of 40 PEO molecules with a polymerization degree of 52 (box size $\sim \rm 5.07 \ nm \times 5.07\ nm \times 5.07\ nm$). The OPLS all-atom force field \cite{Jorgensen1996} is used to describe the PEO molecule, which is generated by the LigParGen web server \cite{Dodda2017} and employs CM1A-LBCC partial atomic charges.\cite{Dodda2017-2} The van der Waals interactions are modeled using the Lennard–Jones potential with a truncation radius of 1.2 nm, consistent with the POPC fine-grained simulation. Electrostatic interactions are handled using the Particle Mesh Ewald (PME) method, also with a cutoff of 1.2 nm.\cite{Essmann1995} Initially, the steepest descent algorithm is employed to remove any unreasonable structures in the initial configuration.\cite{Wardi1988} For all NPT simulations, the target pressure is set to 1.01325 bar. Following this, a 60 ns annealing process is conducted, where the temperature is gradually increased from 298 K to 1000 K and then reduced back to 298 K three times. Subsequently, a 140 ns simulation is performed to ensure the system achieves equilibrium. These two process use the Bussi–Donadio–Parrinello thermostat \cite{Bussi2007} and Berendsen barostat.\cite{ Berendsen1984} Finally, a 200 ns simulation is performed for sampling, with the system maintained at 298 K and 1.01325 bar using Nosé–Hoover thermostat \cite{Nose1984, Hoover1985} and the Parrinello–Rahman semi-isotropic barostat,\cite{Parrinello1981, Nose1983} respectively. Throughout the simulations, the LINCS algorithm \cite{Hess1997} is employed to constrain all bond lengths including hydrogen atoms, allowing for a larger integration time step, i.e., 2 fs.

\subsubsection*{Details of coarse-grained simulation}
The system used for optimization consists of 40 PEO molecules (box size $\sim \rm 5.11 \ nm \times 5.11\ nm \times 5.11\ nm$). To consider the finite size effect on the optimized parameters, three larger systems with box lengths of 10 nm, 20 nm, and 30 nm are also investigated. The MARTINI3 force field \cite{Souza2021} is employed in this section. We perform a 100 ns annealing process where the temperature is gradually increased from 298 K to 500 K and then decreased from 500 K to 298 K, repeating this process three times. A time step of 20 fs is applied in all simulations. Other simulation settings are kept the same as those used in the fine-grained simulation of PEO.

\subsubsection*{Analyses}
NVT simulations on slabs (two-dimensional periodic) in simulation boxes with fixed size of  $\sim \rm 5.11 \ nm \times 5.11\ nm \times 20\ nm$ are applied to calculate the surface tensions of polymer–vacuum interfaces. The surface tension is calculated using the diagonal elements of the pressure tensor:
\begin{equation}
\gamma=\frac{L_z}{2}\left(\left\langle P_{z z}\right\rangle-\frac{\left\langle P_{x x}\right\rangle+\left\langle P_{y y}\right\rangle}{2}\right),
\end{equation}
where \( \left\langle P_{xx} \right\rangle \), \( \left\langle P_{yy} \right\rangle \), and \( \left\langle P_{zz} \right\rangle \) are the average diagonal elements of the pressure tensor.

The glass transition temperature ($T\rm _g$) is calculated by monitoring the specific volume (reciprocal of the density) of the polymer as a function of temperature. The $T\rm _g$ is identified as the temperature at which there is a noticeable change in the slope of the specific volume-temperature curve (Fig. \ref{Fig5}c and S4), indicating a transition from the glassy state to the rubbery state.


\section*{Data availability}
The data that support the findings of this study are available from the authors upon reasonable request. The setup of the examples in this work can be accessed at \url{https://amofms.readthedocs.io/en/latest/example.html}

\section*{Code availability}

AMOFMS is an open-source software package fully developed in Python3, using GROMACS \cite{VanDerSpoel2005, Abraham2015} and MDAnalysis  \cite{Michaud-Agrawal2011} for molecular dynamics simulations and data extraction, respectively. The package is available for easy installation and upgrades through Anaconda Cloud and PiP, ensuring user-friendly accessibility. The source code is managed via GitHub (\url{https://github.com/Dropletsimuli/AMOFMS}), allowing users to download, customize, and contribute to the software's development. Comprehensive documentation (\url{https://amofms.readthedocs.io}) is provided to facilitate usability. \\

\section*{Acknowledgments}
This work was supported by the National Natural Science Foundation of China under Grant No. 22273112 and the National Key R\&D Program of China (No. 2021YFB3803200).
     

\bibliography{sci}

\bibliographystyle{sn}

\end{document}